\documentclass[conference]{IEEEtran}

\makeatletter

\def\ps@IEEEtitlepagestyle{%
    \def\@oddfoot{\mycopyrightnotice}%
    \def\@evenfoot{}%
}
\def\mycopyrightnotice{%
    {\footnotesize  978-1-7281-5242-4/19 \textcopyright2019 IEEE\hfill}
    \gdef\mycopyrightnotice{}
}

\usepackage[]{graphicx}
\usepackage{caption}
\usepackage{subcaption}
\usepackage{float}
\usepackage{fancybox, graphicx}
\usepackage[export]{adjustbox}

\usepackage{eso-pic}
\newcommand\AtPageUpperMyright[1]{\AtPageUpperLeft{%
 \put(\LenToUnit{0.1\paperwidth},\LenToUnit{-1cm}){%
     \parbox{1\textwidth}{\raggedleft\fontsize{10}{5}\selectfont #1}}%
 }}%
\newcommand{\conf}[1]{%
\AddToShipoutPictureBG*{%
\AtPageUpperMyright{#1}
}
}

\begin{document}

\title{Automatic Detection of Satire in Bangla Documents: A CNN Approach Based on Hybrid Feature Extraction Model }

\conf{International Conference on Bangla Speech and Language Processing(ICBSLP), 27-28 September, 2019}

\author{
\IEEEauthorblockN{Arnab Sen Sharma}
\IEEEauthorblockA{Computer Science and Engineering\\
Shahjalal University of\\ Science \& Technology\\
Sylhet-3114, Bangladesh\\
Email: arnab.api@gmail.com}

\and

\IEEEauthorblockN{Maruf Ahmed Mridul}
\IEEEauthorblockA{Computer Science and Engineering\\
Shahjalal University of\\ Science \& Technology\\
Sylhet-3114, Bangladesh\\
Email: mridul-cse@sust.edu}

\and

\IEEEauthorblockN{Md Saiful Islam}
\IEEEauthorblockA{Computer Science and Engineering\\
Shahjalal University of\\ Science \& Technology\\
Sylhet-3114, Bangladesh\\
Email: saiful-cse@sust.edu}
}
\maketitle

\begin{abstract}
Wide spread of satirical news in online communities is an ongoing trend. The nature of satires are so inherently ambiguous that sometimes it's too hard even for humans to understand whether it's actually satire or not. So, research interest has grown in this field.
The purpose of this research is to detect Bangla satirical news spread in online news portals as well as social media. In this paper we propose a hybrid technique for extracting feature from text documents combining \textit{Word2Vec} and \textit{TF-IDF}. Using our proposed feature extraction technique, with standard CNN architecture we could detect whether a Bangla text document is satire or not with an accuracy of more than 96\%.

\end{abstract}


\IEEEpeerreviewmaketitle

\section{Introduction}
Satires can be considered as a literary form which involves a delicate balance between criticism and humor. Through satire or sarcasm, messages are conveyed in an artistic form that sometimes creates a deviated implicit meaning. The goal of satire is not always to tell the truth. Sometimes humans are not effective enough to distinguish between satires and actual news because often the satires are so ambiguous that it is easy to get deceived.

The spread of satirical news is not a new concept. But in the recent years it has become a real threat that can not be ignored anymore.
Easy access of Internet and hyperactivity of users in various social media platforms has given rise to the extensive spread of satirical news. Internet has largely replaced traditional news media. Many people, especially a huge portion of youth depend on Internet and social media as the primary
source for news consumption because of their easy access, low cost and 24/7 availability. They simply believe in what they read in internet and spread the news what they assumed to be true. So, most of the times, satires are not spread with an intention to deceive. But sometimes some people for their personal benefits take advantage and promote the spread of satires as actual news.

As a matter of fact, there are some web based applications such as Snopes.com, FactCheck.org,
PolitiFact etc which act as fact-checkers. But, these services use human staffs to manually check
facts. Though these services provide accurate information most of the time, these are not efficient
enough since they are not automated.
We propose an automated system based on Convolutional Neural Network and Natural Language Processing
to address the problem.

There are some related existing works. The \textit{Literature Review} section will discuss about these. Also, we'll define some terms and techniques that we used in this work.

\subsection{Literature Review}

De  Sarkar, et al. proposed a hierarchical deep neural network approach to detect satirical fake news which is capable of capturing satire both at the sentence level and at the document level \cite{ref20}. Burfoot, et al. used SVM and bag-of-words to detect satires \cite{ref9}. They used binary feature weights and bi-normal separation feature scaling for feature weighting. They got a best overall F-score of 79.8\%\cite{ref9}.

Rubin, et al. classifies news as \textit{Satires, Fabrications and Hoaxing} as the parts of fake news \cite{ref2}. Reyes, et al. used figurative language processing for humour and irony detection \cite{ref10}.

Ahmad and Tanvir used tockenized, stopword free and stemmed data to classify satire and irony using SVM and got an accuracy of 83.41\%\cite{ref11}. el Pilar Salas-Zárate and María used some psycholinguistic approaches for satire detection in twitter and got F-score of 85.5\% for mexican data and 84.0\% for spanish data \cite{ref12}.  

Tacchini and Eugenio check facts using the information of the users who liked a news\cite{ref8}. Applying logistic regression on the information of likers, around 99\% accuracy is achieved for their dataset.

Some approaches simply used a naive bayes classifier to classify a news. After a little bit of preprocessing on 1-grams from the news context, words are fed to a naive bayes classifier. Granik, et al. proposed to do a stemming to increase accuracy\cite{ref5}. The accuracy got using this approach without preprocessing is nearly 70\%\cite{ref5} and Pratiwi, et al. got accuracy of 78.6\% with
preprocessing in Indonesian Language\cite{ref6}.

Ruchansky, et al. proposed a Capture Score and Integrate model\cite{ref7}. Sense making words from the body text of the news of \textit{twitter}\cite{link2} and \textit{weibo}\cite{link3} are taken and fed to a RNN and the reviews of the news are taken as feature. The accuracy for this model is 89.2\% for \textit{twitter} data and 95.3\% for \textit{weibo} data\cite{ref7}.

Conroy,et al. proposed two different approaches for detecting fake news \cite{ref1}. One is linguistic approach which includes deep syntax analysis and semantic analysis. Deep syntax analysis is implemented based on Probability Context Free Grammars(PCFG). It can predict falsehood approximately
91\% accurately.

Another one is network approach that is based on fact checking using the knowledge networks formed by interconnecting the linked data. This approach gives an accuracy
in the range of 61\% to 95\% for different subject areas.

\subsection{Definitions}
\subsubsection{Word Embedding}

Word embedding simply refers to vector representation of words. Normally machine learning models are not capable of processing string or text as input. 
These models expect vectors or values as input. So, transformation of a word to a vector is a crucial part. There are several techniques to convert word
to vectors. These techniques can be categorized as two types
1. Frequency based (TF-IDF, CountVectorizer, HashVectorizer)
2. Prediction based (Word2Vec) 

\subsubsection{TF-IDF}
In this work we focused on TF-IDF vectorizer amongst TF-IDF, CountVectorizer, HashVectorizer etc. as the frequency based word embedder.
TF stands for \textit{Term Frequency} and IDF stands for \textit{Inverse Document Frequency}. It is used in text mining as a weighting factor for features. The 
equation representing the TF-IDF weight of a term t for a particular document \textit{\textbf{d}} (given the whole dataset \textit{\textbf{D}} and the number of documents in the dataset \textit{\textbf{N}}) is 
\begin{center}

\begin{equation}
tf-idf(t,d) = tf(t,d) \times idf(t,D)  
\end{equation}

\[Here,\]\\ 
$tf(t,d) = $ Frequency of \textbf{$t$} in \textbf{$d$}\\
$idf(t,D) = log(\frac{N}{\left| \{d \epsilon D : t \epsilon d\} \right|})$ \cite{link4}
\end{center}
TF is upweighted by the number of times a term occurs in an article. And IDF is downweighted by the number of times a term occurs in the whole 
dataset/corpus. So TF-IDF assigns less significant values to words that generally occurs in most documents such as \textit{is, are, be, to, on ... etc}.

\subsubsection{Word2Vec}
\begin{figure}[H]
  \centering
  \begin{minipage}[b]{0.24\textwidth}
    \includegraphics[width=\textwidth]{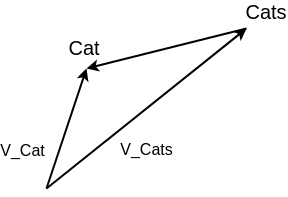}
    \caption{Word2Vec (a)}
    \label{imgRef1}
  \end{minipage}
  \hfill
  \begin{minipage}[b]{0.24\textwidth}
    \includegraphics[width=\textwidth]{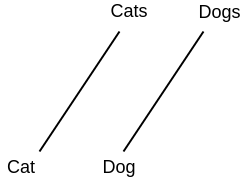}
    \caption{Word2Vec (b)}
    \label{imgRef2}
  \end{minipage}
\end{figure}

Though heavily used in the field of NLP, frequency based word embedders fail to capture the semantic value of a word or document. Word2Vec is the 
process of transforming words to vectors preserving some of their syntactical and semantic correlations. Word2Vec tries to determine the meaning of a word 
and understand its correlation with other words by looking at it’s context. For example, lets take two sentences \textit{"Range Rover is great car"} and \textit{"Range 
Rover is a wonderful vehicle"}, then a well-trained Word2vec should be able map similarities between the words \textit{great} and \textit{wonderful} and the words \textit{car} and \textit{vehicle}. Word2Vec 
uses cosine distance over euclidean distance to measure the similarity or distance between two words. Let's take some pair of singular-plural words like \textit{cat} and \textit{cats}, \textit{dog} and \textit{dogs}.
Here the singular-plural relation between the words cat and cats is represented by the cosine
difference between the two words $V_{cat}$ and $V_{cats}$  (\ref{imgRef1}) is given by the equation below\\

\begin{equation}
cosine(V_{cat} , V_{cats}) = \frac{V_{cat} \times V_{cats}}{||V_{cat}|| \times ||V_{cats}||}    
\end{equation}

And now the word pair dog and dogs have the same singular-plural relationship between them like cat and cats (\ref{imgRef2}). So according to Word2Vec

$$V_{dog} - V_{dogs} = V_{cat} - V_{cats}$$
$$\Rightarrow V_{dogs} = V_{cats} - V_{cat} + V_{dog}$$

So, if we know the particular relationship between two words and know one of the words, Word2vec can predict the other word. 

\subsubsection{CNN in NLP}
Convolutional Neural Network(CNN) is a deep learning algorithm that takes a
multidimentional vector or an image as input. CNNs captures the significant aspects of the input through the application of
appropriate \textit{filters} and perform classification tasks. With enough training CNNs are able to learn which filters are appropriate for different contexts. Different filters/kernels are slided on the beginning layers named after convolutional layer and extract different features
and feed-forward the values to next layer of the architecture. Besides learning high level features of an image these filters also reduce the size of convolved feature 
space and thus reduce the computational power required for processing the data. The convolved features extracted by the convolutional layers are
then fed to a normal neural network architecture possibly with a number of hidden layers. This neural network learns the convolved feature vector and 
does the actual classification task.

Recently CNNs are used heavily in NLP. CNN expects the input to be a multidimensional image but we have a one dimensional vector from a word after word 
embedding. So, instead of single words we feed whole sentences or documents into CNN . From a 20 word document or sentence where each word is embedded to 
a 200 dimension vector, we can get a $20\times200$ matrix. This two dimensional vector can act as an image and can be fed as an input to CNN. From various 
experiments and researches it was found that CNN performs quite well in generalizing the relationships between words in a document and thus capturing it\'s 
semantic meaning. CNN performs better than the simple bag of words and prone to less inaccurate assumptions.

\section{Proposed method}
First of all, we built our own Word2Vec model modifying the traditional Word2Vec model. Then an image is created from a preprocessed document combining Word2Vec and TF-IDF vectorizers. Finally, that image is used as input to a CNN architecture. The detailed 
procedure is discussed below.

\subsection{Building our own Word2Vec model}
There are some great Word2Vec models for English language, but as per our knowledge there is no good performing  Word2Vec model for Bangla. So, we had to create our own Word2Vec model.
To do so, we relied on \textit{gensim} library of python. We needed a lot of textual data to train the model. We used the \textit{scrapy}
library of python to build crawlers and crawled Bangla textual data from Wikipedia and online news portals. We collected 380832 articles in total and used these to train our model. Our Word2Vec model converts Bangla words to a vector of size 10. To check the performance of our model, we checked the 5 most similar words of
a Bangla word. Here are the results.

\begin{figure}[H]
  \begin{center}
     \includegraphics[scale=0.5]{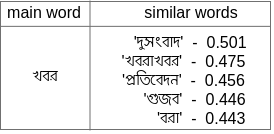}
     \caption{\textit{Testing Word2Vec model with a word}}{}
      \label{}
  \end{center}
\end{figure}

\subsection{Dataset}
We used the \textit{scrapy} library of Pyhon to build crawlers to crawl bangla textual data from different websites. For authentic news data we
crawled news articles from two bangla news portals \textit{Prothom Alo}\cite{link5} and \textit{Ittefaq}\cite{link6}. For satire data we crawled articles from a renowned satire news 
portal \textit{Motikontho}\cite{link7}. We crawled a total of 1480 articles from \textit{Motikontho}. To balance our dataset we randomly selected 1480 articles from the REAL 
news articles we collected from \textit{Prothom Alo} and \textit{Ittefaq}. The formation of the dataset is very  simple. A single data is only a document and a label (satire or not). 

\subsection{Data Preprocessing}
The collected dataset might be mixed with some noisy and unnecessary data. So, we had to get
rid of them through a bit of preprocessing. Our preprocessing consisted of the following steps.
\subsubsection{Ignoring stopwords and punctuations}
We ignored the stopwords from every news document. Because, these words appear in almost
every article and do not provide any significant information. Some example of stopwords -
\begin{figure}[H]
  \begin{center}
     \includegraphics[scale = 0.5]{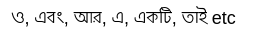}
      \label{}
  \end{center}
\end{figure}

We collected the list of Bangla stopwords from a github repository by \textit{genediazjr}\cite{link8}.

\subsubsection{Stemming}
The purpose of a stemming is to find the root word. We used a stemmer developed by Rafi Kamal\cite{link1} which we found in his github repository. The stemmer performed better compared to other available Bangla stemmers that we could find.

\begin{figure}[H]
  \begin{center}
     \includegraphics[scale=0.6]{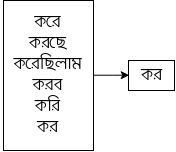}
     \caption{\textit{Stemming}}{}
      \label{}
  \end{center}
\end{figure}

\subsection{Document to Image}
For converting a document to a vector we selected TF-IDF vectorizer of size 1000. It means 1000 words/terms were selected based on our corpus. We ignored the terms that occur in more than 70\% of the documents in our corpus. Because common words do not add any significant value to a specific document. We also ignored rare terms(the terms that occur in less than 10\% of the documents). Because these terms might overfit any model. Also any numeric terms were ignored. The term frequency of the rest of the words were calculated and most significant 1000 terms were selected.   
The respective TF-IDF values of these terms for a document are used to create a vector of size 1000 that represents the document. But these TF-IDF values do not 
hold any semantic meanings. So each of the words were converted to a vector of size 10 using our Word2Vec model. These vectors of size 10 were multiplied
by their respective TF-IDF values for a document. So, for each document we had a 2D vector of size $1000\times10$.
Also, convolutional layers expect pixel values of an image. Pixel values are never negative. So, CNNs cannot take a vector which contains some negative
values. But word2vec embedded vectors can contain negative values. 
CNN input layers actually take a 3 dimensional vector. First two dimensions represents the 2D image and the third dimension is the number of filters. For 
coloured images the number of filters is 3(Red, Green, Blue) and each pixel value of image is formed with 3 values in RGB system. Our strategy to handle negative numbers
was to separate the positive and negative values in two dimensions just like the picture below.

\begin{figure}[H]
  \begin{center}
     \includegraphics[scale=0.40]{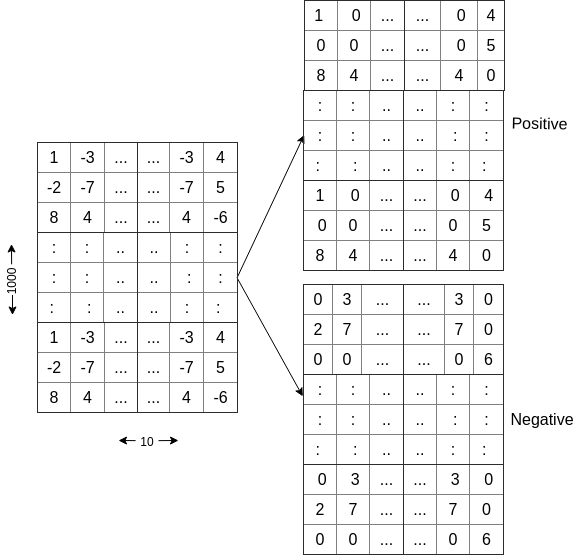}
     \caption{\textit{Transformation of Feature Vectors}}{}
      \label{}
  \end{center}
\end{figure}

So, for each document we had a 3D vector of size $1000\times10\times2$.

\subsection{Structure of the model}
\begin{itemize}
    \item Input layer: Input layer contains a Convolutional2D layer with $256$ filters that takes vectors of shape $1000\times10\times2$ as input.
    \item Another Convolutional2D layer with $128$ filters with \textit{ReLU} activation.
    \item Pooling layer of size $2\times2$.
    \item Dropout layer with value $0.25$ to avoid overfitting.
    \item A Dense layer with $512$ neurons with \textit{ReLU} activation.
    \item Dropout layer with value $0.5$.
    \item Output layer: One neuron with \textit{sigmoid} activation.
\end{itemize}

\section{Results and Analysis}
The dataset was randomly split to two parts. 70\% of the data (2018 documents) were used as training dataset. The rest 30\% (942 documents) were used to test
the performance of our model. This model gave us an accuracy rate of 96.4\% on the test dataset. Since the dataset was balanced the F1 score was same as the accuracy value. The confusion matrix is given below.

\begin{figure}[H]
  \centering
  \begin{minipage}[b]{0.25\textwidth}
    \includegraphics[width=5cm, height=5cm]{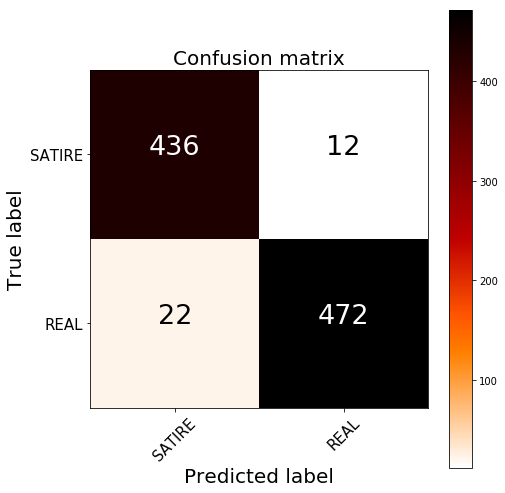}
    \captionsetup{labelformat=empty}
    \caption{(a) Total Count Representation }
  \end{minipage}
  \hfill
  \begin{minipage}[b]{0.25\textwidth}
    \includegraphics[width=5cm, height=5cm]{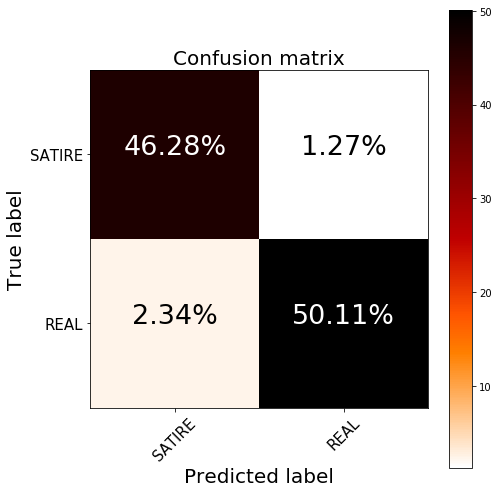}
    \captionsetup{labelformat=empty}
    \caption{(b) Percentile Representation}
  \end{minipage}
  \caption{\textit{Confusion Matrices showing the results}}
\end{figure}

There are some scopes to improve our Word2Vec model  and a perfect stemmer for Bangla language can boost the overall performance of our proposed model. The model was compiled with 2GB graphics card of NVIDIA GeForce 940M. So,
we could not use a TF-IDF vector and a Word2Vec vector of larger size. If we had access to more resources we could use bigger feature vectors and the
accuracy might have improved some.

Actually, in terms of accuracy, humans are much more effective than a machine for this task. Maybe for our dataset, human will be able to detect satires 100\% accurately. But, though the accuracy falls a bit, we think it's much better to use an automated approach which saves a lot of time.

\section{Conclusion}
Satire detection for Bangla news is completely new. As per our knowledge, no such work has been done in this sector for Bangla language. We found that our hybrid feature extraction technique combined with a  CNN model performs great in language processing for pattern finding.
Since satire is a type of fakeness, satire detection can be an important prerequisite of fake news detection. So, this work might be helpful to take better decisions on fake news detection and other such works. Also our hybrid feature extraction technique can be used in other works of similar nature.

\end{document}